\documentclass[10pt,letterpaper]{article}
\usepackage{opex3}

\def\aap{A\&A\ }
\def\mnras{Mon. Not. R. Astron. Soc.\ }
\def\aaps{A\&AS\ }%
\def\pasp{Astron. Soc. Pac.\ }%

\begin{document}

\title{Pupil remapping for high contrast astronomy: \\ 
  results from an optical testbed}

\author{T. Kotani$^1$, S. Lacour$^{1,2*}$, G. Perrin$^1$, G. Robertson$^2$ and P. Tuthill$^2$} 

\address{$^1$LESIA, Observatoire de Paris, CNRS/UMR 8109, 92190 Meudon, France\\
$^2$School of Physics, Sydney University, NSW 2006, Australia}

\email{$^*$Corresponding author: sylvestre.lacour@obspm.fr} 



\begin{abstract}
The direct imaging and characterization of Earth-like planets is among
the most sought-after prizes in contemporary astrophysics, however
current optical instrumentation delivers insufficient dynamic range to
overcome the vast contrast differential between the planet and its
host star. New opportunities are offered by coherent single mode
fibers, whose technological development has been motivated by the
needs of the telecom industry in the near infrared. This paper
presents a new vision for an instrument using coherent waveguides to
remap the pupil geometry of the telescope. It would (i) inject the
full pupil of the telescope into an array of single mode fibers, (ii)
rearrange the pupil so fringes can be accurately measured, and (iii)
permit image reconstruction so that atmospheric blurring can be
totally removed. Here we present a laboratory experiment whose goal
was to validate the theoretical concepts underpinning our proposed
method. We successfully confirmed that we can retrieve the image of a
simulated astrophysical object (in this case a binary star) though a
pupil remapping instrument using single mode fibers.
\end{abstract}

\ocis{(350.1260) Astronomical optics; (060.2430)   Fibers, single-mode; (070.6110)   Spatial filtering; (120.3180)   Interferometry.} 

\section{Introduction}

Since the discovery of a giant planet around the solar-type star 51
Pegasi by Mayor \& Queloz\cite{1995Natur.378..355M}, over 200
extrasolar planets have been discovered, almost all by indirect
methods such as radial velocity monitoring and transit
observations. Selection biases inherent to these techniques helps
explain the high incidence of relatively massive (Jupiter-size)
planets orbiting close to their parent stars in the sample so far. At
visible wavelengths, the reflected light from such an extrasolar giant
planet is typically more than $10^4$ times fainter than the direct
light from its parent star\cite{2000ApJ...540..504S}.  Bright speckles
arising naturally from the transfer function of the turbulent
atmosphere are a good mimic of the expected planetary signal,
compounding the difficulty of the direct imaging problem.  A plethora
of technologies have been advanced to tackle the high-contrast imaging
problem, however it is fair to say that none so far has achieved wide
success except over quite narrowly-defined niches, as discussed
further below.

By partially correcting the atmospheric degradation, Adaptive Optics
(AO) has led to several direct detections of extrasolar planets
\cite{2004A&A...425L..29C,2008arXiv0811.2606M}. However, the detection
limit is highly dependant on the angular
separation\cite{2005ApJ...629..592G}.  Published limits are, up to
now, of the order of $10^3$ at 500 milliarcseconds, and $10^4$ at
twice that distance\cite{2008arXiv0811.3583L}.  At smaller separations
(1 to $\approx 5$ $\lambda/D$), Fourier transform deconvolution
techniques is a necessary complement to AO imaging. In that domain,
speckle imaging\cite{1970A&A.....6...85L} and aperture masking
interferometry\cite{1987Natur.328..694H,2000PASP..112..555T} that
utilize post-processing of rapid-exposure data have demonstrated
recovery of high contrast images.  In particular, masking sets the
present standard for very high angular resolution work within a few
tenths of an arcsecond, although over such scales the best
demonstrated dynamic range does not exceed a few hundred. Last but not
least, long-baseline optical interferometry also has a role in faint
companion detection, with a dynamic range of a few $10^3$ at angular
separation of a few tens of milliarcseconds\cite{2008A&A...485..561L}.

The limited dynamic ranges attained by these techniques can mostly be
attributed to the fact that atmospheric turbulence is highly
time-variable, resulting in difficulty with the wavefront calibration.
By tackling this problem, a single-mode pupil remapping
system\cite{2006MNRAS.373..747P} is, on the paper, an attractive
solution to achieve a high dynamic range (up to
$10^6$)\cite{2007MNRAS.374..832L} at the resolution limit of a
telescope. The advantages of a single-mode pupil remapping system are
as follows: first, the use of single-mode fibers filters out
atmospheric turbulence effects, as already demonstrated by single-mode
fiber long-baseline interferometry\cite{1997A&AS..121..379C}.  Second,
the non-redundant pupil configuration eliminates redundancy noise
which affects the accuracy of wavefront measurements. Third, the full
pupil area of the primary mirror can be used thanks to the pupil
remapping by fibers. The removal of atmospheric turbulence and
redundancy noise allows almost perfect calibration of the degraded
wavefront. Therefore this technique can take full advantage of the
intrinsic high angular resolution of large ground-based telescopes and
fully realize the aperture-related gains over and above their large
photon-collecting capability.

The technique described here originally grew out from ideas for
improvements to the successful aperture masking experiments which are
now established at many major world observatories (e.g. Keck, Gemini,
VLT).  This historical legacy gives us a straightforward model to
assist in constructing the arguments in favor of pupil remapping, and
we draw upon this earlier work for examples. Following a brief review
of the principles in Section~\ref{sc:Op}, this paper presents a practical
laboratory experiment which has validated the principles of the
proposed instrument. We show that image reconstruction is possible,
even though the pupil geometry was considerably altered.  Details of
the optical design for the testbed are given in Section~\ref{sc:pro}.  Results
and conclusions are given in the final section.

\section{Operational Principles}\label{sc:Op}

\subsection{Aperture Masking}

\begin{figure}
\centering\includegraphics[width=4cm]{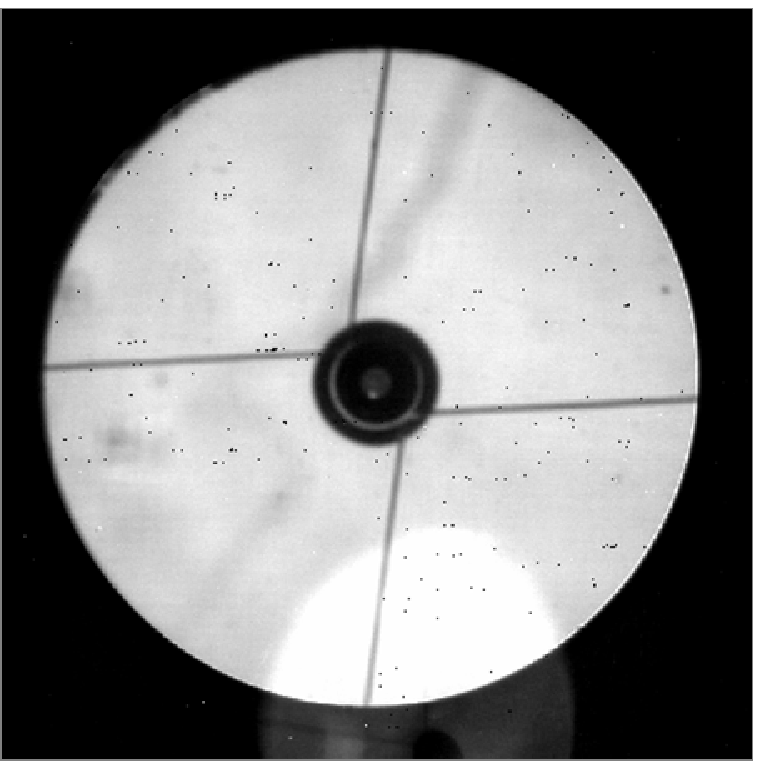}
\centering\includegraphics[width=4cm]{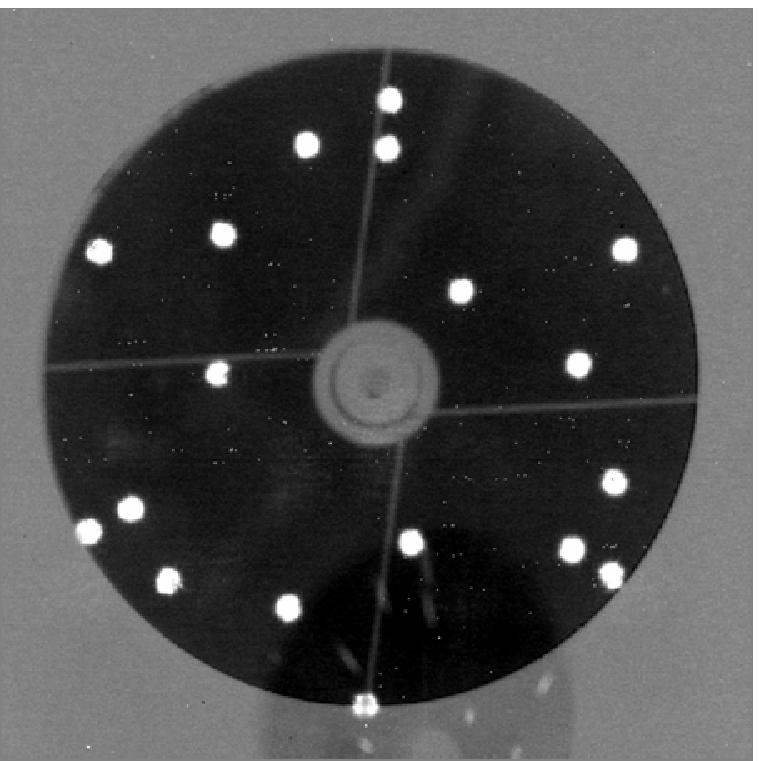}
\centering\includegraphics[width=9cm]{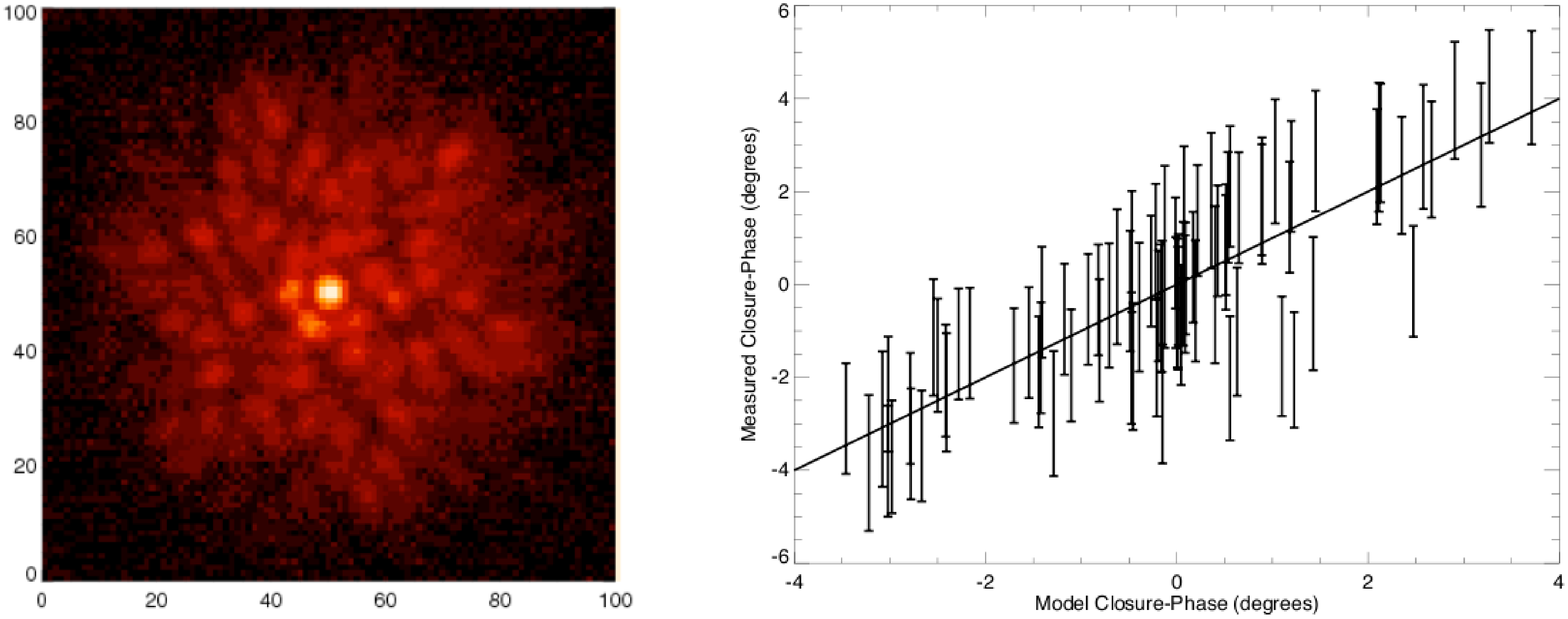}
\caption{A pictorial description of the aperture masking experiment
  now commissioned at the NAOS/CONICA camera on one of ESO's four VLT
  telescopes in Paranal. Top Left Panel: the unobstructed telescope
  pupil complete with spiders and central obstruction, as viewed in
  the CONICA camera. Top Right: an 18-hole aperture mask transforms
  the 8\,m primary mirror into a sparse, non-redundant array. Note the
  non-ideal mask alignment visible in these engineering-test
  images. Lower Left: the diffraction pattern created in the image
  plane after using such a non-redundant pupil mask. With the adaptive
  optics system operating, a bright core persists at the center of the
  pattern, which otherwise spreads light over an area in inverse
  proportion to the size of the holes in the mask plate. Lower Right:
  this panel depicts the final data product for faint companion
  searches -- the closure phase recovered by the data reduction
  software over all closed triangles of holes in the mask. For a
  simple unresolved point, this data product should give a value of
  zero (to within the limits of the random noise level).  Here we see
  a systematic variation of the closure phase. When plotted against a
  model of two point-like sources, the variation betrays the presence of
  the high-contrast companion.
\label{fig1}}
\end{figure}

Very high dynamic range imaging requires the minimization of the
number and intensity of speckles in the image. Adaptive optics has
been pursued as the most conceptually straightforward (although
technically demanding) solution. At visible wavelengths, a deformable
mirror with a very large number of actuators is required. This is now
feasible for a 4\,m telescope, although there are limitations and high
dynamic range very near to the central PSF is generally not
obtainable. Of order 200 actuators are required to achieve basic
diffraction limited imaging. However the high dynamic range
performance needed for Jupiter mass planetary detection requires at
least several thousands actuators in the visible, together with a
high temporal sampling frequency of the wavefront. These requirements
drive the adaptive optics system into technical difficulties and high
cost.

An alternative is data post-processing.  The family of techniques
involved is mainly known as speckle imaging or speckle interferometry
\cite{1970A&A.....6...85L}.  The fundamental idea is that blurring from
atmospheric turbulence can be frozen using exposures shorter than the 
seeing coherence time, allowing turbulence-free spatial information to 
be recovered during post-processing of the data. The post-processing is 
performed in the spatial frequency plane, the plane conjugated to the sky 
plane by the Fourier transform. Speckle imaging proceeds by calibrating 
the energy at each spatial frequency thanks to an identical observation 
of an unresolved star used as a reference. If the statistical properties 
of the turbulence were similar for the reference star and science target 
then some degree of correction can be made and the image partially cleaned
of seeing-related artifacts by this procedure. Unfortunately atmospheric 
turbulence is not stationary and rapid changes of the transfer function
means that even this statistical calibration is imperfect. 

One key shortcoming of the method is that multiple different areas within
the pupil telescope all contribute to the same spatial frequency in the
image plane (this is called redundancy). Because different patches 
sum together with essentially random phases, the energy recorded at a 
given spatial frequency is both reduced and contaminated with seeing noise. 
Since decorrelation between phases increases with the distance between
patches in the pupil plane, two different sets of patches (or
sub-pupils) corresponding to a same spatial frequency will have less
and less correlated phases leading to a suppression of the transfer
function at increasing spatial frequencies: high spatial frequencies
are more blurred out than low ones. These effects are all
detrimental to high dynamic range imaging.

Closely related to speckle imaging is the technique of aperture
masking \cite{1987Natur.328..694H}. The principles are similar to
speckle imaging, but with a mask placed in a pupil plane to select
only non-redundant sub-apertures. Spatial frequencies therefore
correspond to unique pairs of sub-apertures from the pupil plane. As a
consequence, high spatial frequencies are no longer attenuated with
respect to lower spatial frequencies and the dynamic range in
reconstructed images is the same at all scales. This technique has
been successfully used on several telescopes. As an example,
Fig.~\ref{fig1} is a demonstration of faint companion detection
obtained with the VLT at near-infrared wavelengths. This technique
still has some drawbacks as far as dynamic range is
concerned. Firstly, only a small fraction of the pupil is used: each
sub-aperture has a size of the order of the Fried parameter
$r_0$. Because of the non-redundancy requirement, only a limited
number of sub-apertures are employed. The total pupil area used to
collect photons (3\% in the case of this VLT experiment) is
illustrated by the pupil imaged in the upper panels of
Fig.~\ref{fig1}.  Second, the size of sub-pupils is such that the
phase varies slightly across them, inducing some
decorrelation. Because the system is therefore not completely immune
to phase fluctuations, some residual atmospheric turbulence noise
persists in the data collected, placing limits on the dynamic range
obtained.

\subsection{Aperture masking with single-mode fibers}

\begin{figure}
\centering\includegraphics[width=13cm]{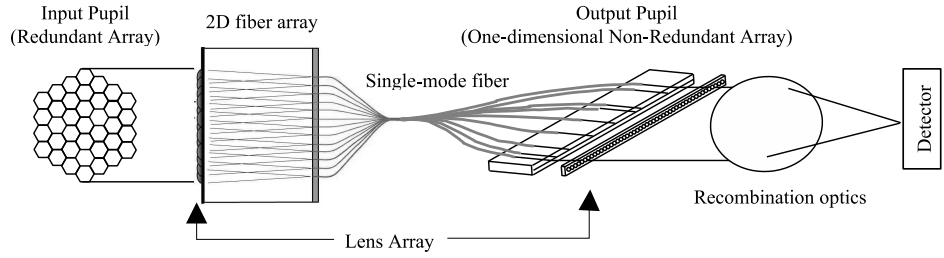}
\caption{A schematic diagram illustrating the principles of pupil
  remapping with single mode fibers. In brief, the left side depicts
  the pupil fragmentation and single-mode fiber injection optics,
  while to the right we show the spatially-filtered beams output from
  the fibers into a custom-designed non-redundant linear array.  These
  elements are all described in some detail in our earlier
  papers\cite{2006MNRAS.373..747P,2007MNRAS.374..832L} 
\label{fig2}
}
\end{figure}

The problems of aperture masking can be solved by using single-mode
fibers.  First of all, single-mode fibers are perfect spatial filters:
any light injected into a single-mode fiber will emerge as a Gaussian
beam with a flat wavefront. As a consequence, wavefronts will be fully 
coherent after filtering with single-mode fibers. The cost of this 
mode cleansing is that coupling of an imperfect and variable beam into 
the fiber core can be low, and highly time-variable. But the final 
result is that the wavefront can be considered ideally flat over the
entire fiber input aperture, and when separate beams filtered in this
way are interfered, the data will be uncorrupted by phase corrugations.
Consequently, an important source of dynamic range loss, speckle
noise, is eliminated.  Second, fibers are flexible optics and it is
possible to use the full pupil in the input and redistribute the
sub-pupils in the output in a non-redundant way. The full photon
collecting capability of the telescope is therefore available for
interferometric operation. The principles of the system are depicted 
in the sketch Fig.~\ref{fig2}.

The instrument therefore acts to turn the telescope into a multiple-beam
interferometer. Each pair of fibers maps to a specific baseline in the
telescope pupil, for which a complex visibility can be measured and 
calibrated. An image can then be reconstructed from the full
set of such visibilities by exploiting Fourier techniques already 
well-established across a broad range of astronomical interferometers.

\section{A laboratory demonstrator} \label{sc:pro}

\subsection{Overview}

\begin{figure}
\centering\includegraphics[height=8cm]{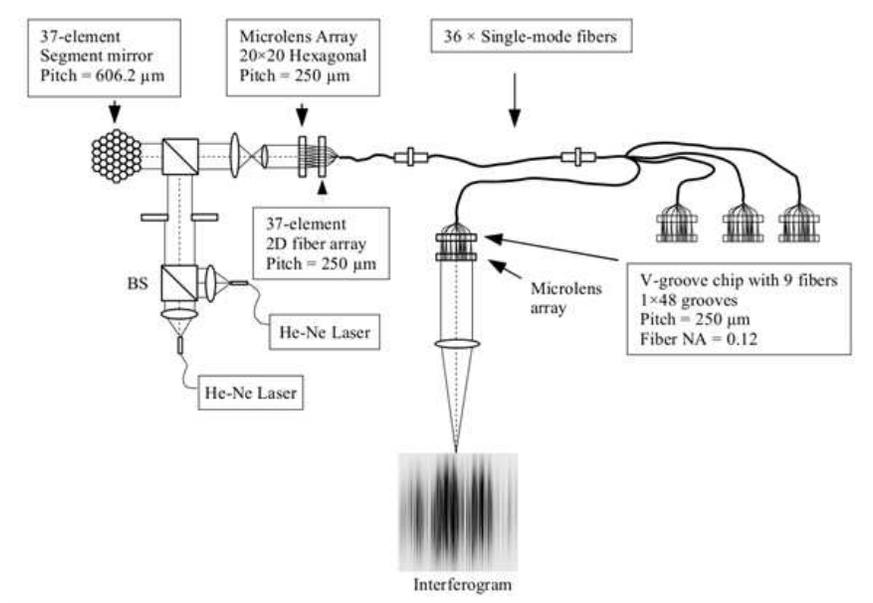}
\caption{Experimental setup. The optical layout of a recombiner is shown
for only one v-groove chip. The other 3 v-groove chips may share the same
recombination optics. Further details of the various components and subsystems
depicted can be found in the text.}
\label{Sch}
\end{figure}

Two instrument concepts are under development, with the ultimate aim
of providing a versatile instrument to the astronomical community. One
of them called FIRST (Fibered Imager foR Single Telescope) is being
developed at Paris Observatory as a phase A study of the new
generation of instruments to equip the CFHT (Canada-France-Hawaii
Telescope) by mid-2013.  The second instrument is called Dragonfly and
is under development at The University of Sydney, for use at the
Palomar 5m telescope as a visitor instrument.

The two instruments are similar in concept and required the setup of an optical
testbed. A simple schematic for the FIRST testbed is shown in Fig.~\ref{Sch}, 
which has followed a number of key design principles and components:  
\begin{itemize}
\item Operating wavelength: 600nm to 800nm. This wavelength range is set by the
cut-off wavelength of a single-mode fiber ($\approx 550$ nm) and the
sensitivity of the detector.  
\item Segmented Deformable Mirror is used for precise fiber coupling.  
\item Two-dimensional fiber array and a microlens array: the input
  pupil of the telescope is fragmented by a microlens array then
  injected into a single-mode fiber array.
\item Spectrometer: the output beams from fibers are spectrally dispersed
to minimize the effect of OPD and dispersion between fibers. Spectral resolution
$R \approx 100$ is required.
\item Beam combiner: the output beams are recombined to measure complex 
visibilities of the target object.
\end{itemize}

To demonstrate the imaging capability of this instrument, an
artificial binary star was simulated for this experiment. Two laser
sources were injected into $5 \mu$m pinholes to make spatially
coherent point sources, and were collimated before injection into the
system. One source was aligned on-axis (simulating the primary star)
and the other was slightly tilted simulating an off-axis source (the
companion). The two beams were then combined using a beamsplitter. A
37-element MEM, segmented deformable mirror (DM) was placed in the
collimated beam to align the tip-tilt of each sub-aperture for precise
injection into fibers. A second beam splitter was used to get
perpendicular incidence on the DM, with a 50\% return loss to the
beamsplitter (the setup would not be identical on a practical
instrument). After reflection by the DM, the beam diameter was reduced
by a factor of 2.4, matching the pitch of adjacent sub-apertures to
that of the single-mode fibers. A microlens array divided the input
pupil into 36 sub-apertures, with each beam being focused onto the
core of an individual fiber in the two-dimensional single-mode fiber
array. This array consists of a bundle of 36 single-mode fibers
precisely aligned on a hexagonal grid.

Once the light is coupled into the fibers, many of the key processing steps
are then naturally achieved. These are the spatial filtering which happens
in fiber, and the pupil remapping which can be trivially achieved by ensuring
that the output pupil differs from the input pupil. For our testbed setup, this
output fiber array took the form of a linear non-redundant array. To construct
such a setup, fibers were precisely aligned in a silicon v-groove chip and the 
output beams from the fibers were recollimated by a linear microlens array. 
The beams were subsequently recombined in the image plane to make a simple 
Fizeau type interference pattern.

\subsection{Injection optics}

\begin{figure}
\centering\includegraphics[height=4cm]{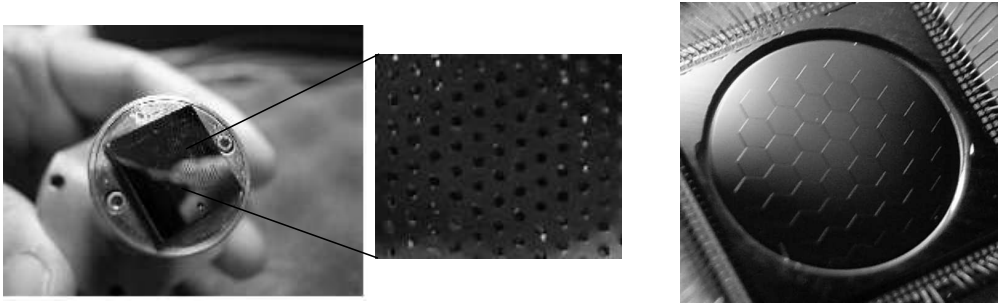}
\caption{Left: 2D fiber array from FiberGuide Industries. 
The fiber pitch is 250 $\mu$m. 
Right: Segmented deformable mirror from IRIS AO. 
The pitch between adjacent segments is 606.2 $\mu$m including a 4 $\mu$m gap. }
\label{fiberA}
\end{figure}

One of the most critical tasks for successful operation of this
system is to realize an extremely high precision optical alignment of the
bundle of 36 single-mode fibers. For optimal coupling, the position of the
fiber cores must be adjusted with sub-micron precision. For this, we
employed a combination of a two-dimensional fiber array, microlens array,
and a segmented deformable mirror. The 2D single-mode fiber array consists of
a bundle of 36 single-mode fibers fabricated by Fiber Guide Industries
(Fig.~\ref{fiberA}). The fibers are packed in a hexagonal arrangement 
within a silicon substrate. The fiber pitch is 250 $\mu$m, while typical 
positional errors were $\approx 1 \mu$m and the angular precision of fiber 
alignment was 2.5 mrad. The single-mode fibers used for the 2D fiber array 
were connectorized by Oz optics. The fiber is a polarization maintaining 
fiber (PM-640HP) optimized for operation in the R band with the following
parameters: diameters of the fiber core/cladding are 4 and 125 $\mu$m
respectively; cutoff wavelength is 550 nm; numerical aperture is 0.12. The
alignment accuracy of the slow axis of the polarization state in the fibers
is $\pm$ 3 degrees. However the precision of the fiber alignment is not
sufficient to achieve very high coupling efficiency and very high dynamic
range. Thus we used a 37-element segmented deformable mirror based on
microelectromechanical systems technology (IRIS AO) for very precise beam
alignment with respect to fiber cores. Each segment consists of a hexagonal
mirror attached to three individual electrostatic actuators, which provide
tip-tilt and piston motion through differential actuation. An actuator has
a 5 $\mu$m stroke. The largest inscribed aperture is 3.5 mm and the
center-to-center distance between two adjacent segments is 606.2 $\mu$m,
including a 4 $\mu$m gap. This DM allows compensation of fiber position
errors in the 2D fiber array. In the demonstration described here, the DM
is used for static positional error compensation only, but in future it may
be possible to correct the tip-tilt and piston errors arising from atmospheric
turbulence using this electro-optical device.

\subsection{Beam combiner, Anamorphic optics and Spectrometer}

\begin{figure}
\centering\includegraphics[height=4cm]{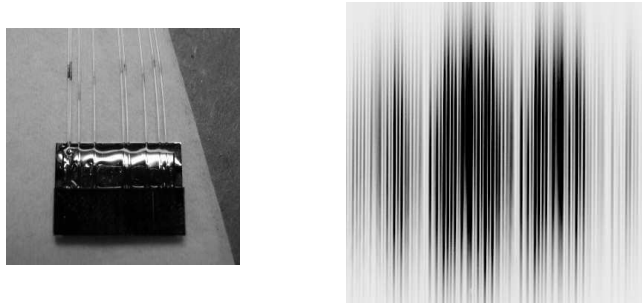}
\caption{Left Panel: the output linear fiber array arranged in the
  silicon V-groove. Right: the resultant intensity distribution
  displaying linear Fizeau interference fringes.  Anamorphic optic is
  later on used to vertically collapse the fringe pattern, allowing
  spectral dispersion. }
\label{vgroove}
\end{figure}

In order to measure fringe visibilities, output beams from the fibers
need to be interferometrically recombined. We employed an image-plane
combination scheme for this instrument in which fibers were arranged
in a non-redundant one-dimensional array on a silicon V-groove chip
from OZ optics (see Fig.~\ref{vgroove}). Such devices are widely used
for telecommunication. The uniform spacing of grooves
($250\pm0.5\,\mu$m) gives precise alignment between fibers. This
device is also used for the MIRC
beamcombiner\cite{2006SPIE.6268E..55M} at the CHARA
interferometer\cite{2005ApJ...628..439M} at Mt. Wilson. The beams from
the fibers are recollimated by a linear microlens array whose lens
pitch is $250\pm0.5\,\mu$m (48 element linear array, SUSS
MicroOptics). Then, they are recombined in the image plane to measure
the object complex fringe visibilities.  The non-redundant
configuration is important so that any fiber pair has a unique
spacing. Unfortunately, due to the limited number of detector pixels,
a non-redundant, 36-element one-dimensional array is difficult to
realize. To achieve sufficient fringe sampling, at least 4 pixels are
needed for the highest frequency fringes to avoid aliasing problems
and significant visibility loss. In the case of a 36-beam combiner,
the required number of pixels is over 10000. Therefore we divided the
36 fibers into 4 sets of 9-beam combiners, with each group of 9 using
one v-groove chip. For a 9-beam combiner, a 48 channel v-groove chip
is used and the fibers are placed on the following positions to ensure
non-redundancy: 2,3,7,14,27,29,37,43,46. The number of fringes inside
the first dark ring in the PSF is about 110, which requires 440 pixels
for sufficient fringe sampling. Eventually, the 4 v-groove chips may
share the same set of recombination optics.

For white light operation using an incoherent source, the optical
pathlength through each fiber pathway in the interferometer must be
matched, as must the amount of glass pathlength.  In order to relax
the strict tolerance on these matching conditions, the recombined beam
is spectrally dispersed by a prism to minimize the effect of residual
Optical Path Differences (OPD) between the fibers. Because the beams
are recombined one-dimensionally, the orthogonal axis can be used to
disperse the light. With sufficiently high dispersion, spectral
channels for the coherent fringe detection are relatively small
leading to long coherence lengths.  Along the dispersion direction,
the diameter of the Airy pattern corresponding to one spectral
resolution element must be small enough to avoid losing spectral
information, while maintaining adequate sampling in the fringe
direction.  The system therefore requires anamorphic optics to give a
differing image scale in the spatial and spectral directions. For a
9-beam combiner, the optimum diameter ratio (anamorphic ratio) is
440:1. Our design for an anamorphic system is relatively simple, using
an afocal combination of two cylindrical lenses with the ratio of
their focal lengths corresponding to a magnification of a pupil
diameter. This results in a small PSF in the spectral dimension in the
image plane. To further increase the anamorphic ratio, we added one
cylindrical lens having a long focal length and having a focusing
power perpendicular to the other two lenses. Zemax simulations taking
into account physical optics propagation showed that the maximum
anamorphic ratio of this system is 130:1. Strong spherical and
chromatic aberrations from the microlens prevent the system from
reaching the optimum anamorphic ratio.
 
Integration times for the camera must be chosen so that atmospheric
piston fluctuations are frozen during an acquisition, thereby keeping
the fringe contrast high. The typical atmospheric coherence time at
visible wavelengths is a few milliseconds. Therefore a key requirement
for the detector is very low read-out noise at very high frame rates,
enabling precise fringe visibility measurements to be obtained. The
EMCCD technology \cite{2001SPIE.4306..289M,2008A&A...480..589B} is a
cost-effective solution for this purpose. Our EMCCD camera, Andor
technology Luca-S, can be read at high frame-rates ($\approx 37$
frames/sec) for reading out a full area (658x496 pixel) and
effectively very small read-out noise ($<0.1$ e$^-$). These features
allow this detector to be used in the photon noise limited regime.

\section{Results and Conclusions}

\begin{figure}{
\centering\includegraphics[height=4cm]{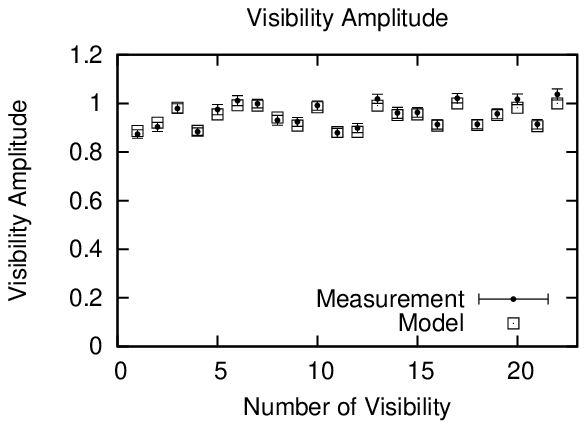}
\centering\includegraphics[height=4cm]{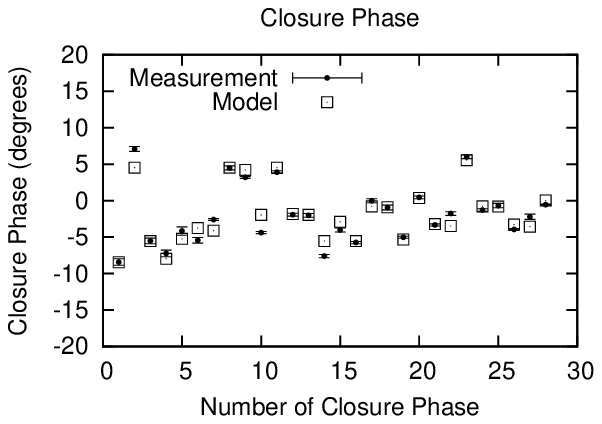} 
\vspace{.2cm}}
\centering\includegraphics[height=4cm]{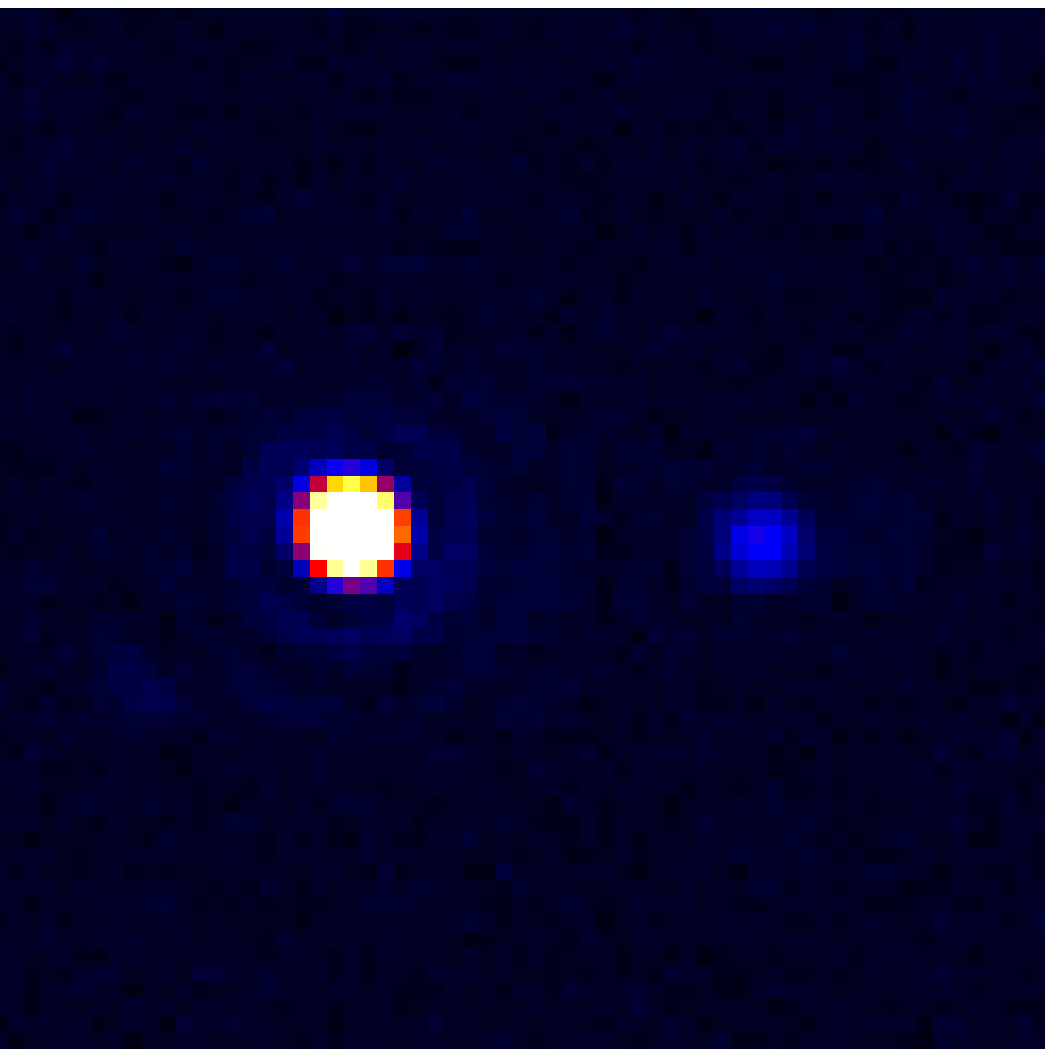} \hspace{1cm}
\centering\includegraphics[height=4cm]{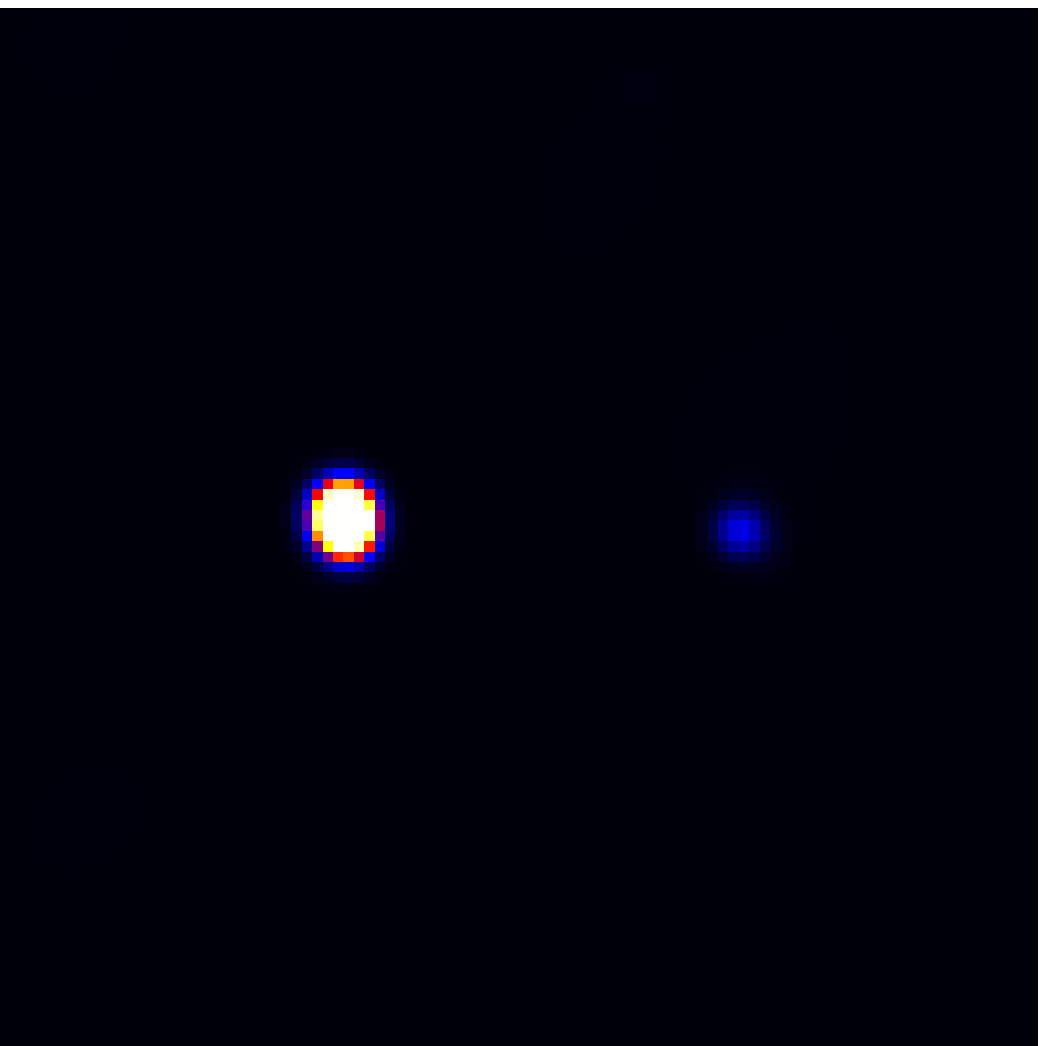}
\caption{Upper left panel shows measured visibility amplitudes,
  together with visibility predictions from the best-fit model for the
  simulated binary star system. Visibility measurement errors were
  found to be about 2\%. Upper right panel shows measured closure
  phases and predictions from the best-fit model. The closure phase
  measurement error was about 0.2 degrees.  Lower Left: Direct image
  of the simulated binary-star source. Right: Reconstructed image from
  complex visibility data recovered from the testbed. A byproduct of
  the image reconstruction algorithm is to deconvolve the point spread
  function, effectively removing on the reconstructed image the Airy
  rings caused by diffraction.}
\label{results}
\end{figure}

The goal of this experiment was to demonstrate real recovery of
diffraction-limited structure from an optical testbed simulating the
FIRST/Dragonfly experimental methodology. Our setup for these
laboratory tests is shown in Fig.~\ref{Sch}.  Although the final
system should use all 36 fibers, for simplicity only 9 fibers were
used for this experiment.  Each input beam from a sub-aperture was
precisely aligned with respect to a fiber for optimum fiber coupling
by using the DM. Once the beams were aligned, the DM kept the mirror
positions fixed during the measurements. We measured fringes from the
artificial binary star generated by He-Ne laser illumination. The
contrast ratio between the two sources was 15. Slow phase fluctuations
were observed, probably thermally induced in single-mode fibers. 100
frames of image data were collected, with an integration time for each
frame of 0.1 second. We fitted a binary star model to the measured
visibilities. The upper panels of Fig.~\ref{results} show measured
average fringe visibilities, closure phases and predictions from the
best fit models. Visibility and closure phase measurement accuracies
were 2\% and 0.2 degrees respectively. The best-fit model is in good
agreement with the measured data. A contrast ratio of $15\pm1$ derived
from the fitting matched well the directly measured value from a
monitor CCD. The separation between the primary and the secondary
source was 0.56 $\lambda/d$, where $d$ is a sub-aperture diameter.

In addition to the direct model fitting to the data, we reconstructed
the image of the simulated binary `star', using the image
reconstruction software MIRA \cite[Multi-Aperture Image Reconstruction
  Algorithm]{2008SPIE.Thiebaut}.  The lower panels of
Fig.~\ref{results} show the reconstructed image (right) and the
directly measured CCD image (left). Two point-like sources are clearly
visible in our reconstructed image, which shows excellent agreement
with the true test object yielding the same flux ratio and separation.

Our experiments have demonstrated the good data quality and the image
reconstruction capability of a pupil remapping system using
single-mode fibers for a single telescope. Thanks to the segmented
deformable mirror, each input beam from a sub-aperture can be aligned
very precisely with respect to a fiber, allowing excellent fiber
coupling. The fringe visibilities obtained from the artificial binary
star in the laboratory were in good accord with expectations from the
theoretical model. Closure phases have been measured with 0.2 degree
stability and fringe visibilities with 2\% accuracy.  Moreover, we
successfully reconstructed the original image from the visibilities
using the MIRA image reconstruction algorithm. This experiment is
therefore an important step toward realizing an on-sky operational
system.


\end{document}